\pdfoutput=1
\documentclass[twocolumn,showpacs,amssymb,floats,prb,superscriptaddress,aps,amsmath]{revtex4}

\pdfoutput=1
\usepackage{graphicx}
\usepackage{dcolumn}
\usepackage{bm}
\usepackage{textcomp}

\begin{document}

\title{Adsorbate-induced Restructuring of Pb mesas Grown on Vicinal Si(111) in the Quantum Regime}

\author{Alexander Ako Khajetoorians}
\affiliation{%
Department of Physics, The University of Texas at Austin, Austin, Texas, 78712 USA}
 \author{Wenguang Zhu}
 \affiliation{Department of Physics and Astronomy, The University of Tennessee, Knoxville, TN 37996 USA}
\affiliation{Materials Science and Technology Division, Oak Ridge National Laboratory, Oak Ridge, TN 37831 USA}
 \author{Jisun Kim}
\affiliation{%
Department of Physics, The University of Texas at Austin, Austin, Texas, 78712 USA}
 \author{Shengyong Qin}
\affiliation{%
Department of Physics, The University of Texas at Austin, Austin, Texas, 78712 USA}
 \author{Holger Eisele}
 \affiliation{%
 Institut f\"{u}r Festk\"{o}rperphysik, Technische Universit\"{a}t Berlin, 10623 Berlin, Germany}
 \author{Zhenyu Zhang}
\affiliation{Materials Science and Technology Division, Oak Ridge National Laboratory, Oak Ridge, TN 37831 USA}
\affiliation{Department of Physics and Astronomy, The University of Tennessee, Knoxville, TN 37996 USA}
\author{Chih-Kang Shih}%
 \email{shih@physics.utexas.edu}
\affiliation{%
Department of Physics, The University of Texas at Austin, Austin, Texas, 78712 USA}

\begin{abstract}
Using scanning tunneling microscopy and spectroscopy, we demonstrate that the adsorption of a minute amount of Cs on a Pb mesa grown in the quantum regime can induce dramatic morphological changes of the mesa, characterized by the appearance of populous monatomic-layer-high Pb nano-islands on top of the mesa. The edges of the Pb nano-islands are decorated with Cs adatoms, and the nano-islands preferentially nucleate and grow on the quantum mechanically unstable regions of the mesa. Furthermore, first-principles calculations within density functional theory show that the Pb atoms forming these nano-islands were expelled by the adsorbed Cs atoms via a kinetically accessible place exchange process when the Cs atoms alloyed into the top layer of the Pb mesa.
\end{abstract}

\maketitle

\section{Introduction}

Thin film technology has greatly advanced over the last few decades and plays an important role in advancing electronic and spintronic-based technology.  On the fundamental side, efforts have been focused on controlling the formation of novel nanostructures through manipulation of strain energy, adsorbate-modified kinetics, and light irradiation \cite{Brune1998,Ernst1998,Peale1992,Ross1999,Stangl2004}.  For ultra-thin metal epitaxy, emerging research over the last decade has demonstrated that quantum confinement of electronic states can have profound effects on the growth of various metal nanostructures as well as their physical and chemical properties \cite{Aballe2004,Eom2006,Evans1993,Guo2004,Hinch1989,Miller1988,Ozer2007,Smith1996}.  This is commonly referred to as ``quantum'' or ``electronic growth \cite{Zhang1998}.''\\
\nolinebreak
\indent Among various systems exhibiting quantum growth behavior, Pb on Si$(111)$ is probably the most thoroughly investigated.  This is due to a nearly 4:3 matching between the $(111)$ lattice constant and the Fermi wavelength $(E_{F}$), resulting in a bi-layer quantum oscillation of the density of states \cite{Wei2002,Jia2006}.  This thickness-dependent modulation of the quantum well states (QWS) has been utilized to study the interplay between quantum confinement and surface diffusion of Fe adatoms as well as QSE-induced oscillations of surface reactivity for adsorbed O atoms \cite{Ma2006,Ma2007}.  However, for these specific cases, there is no experimental evidence showing that adsorbates induce modifications to the underlying quantum growth phenomena. Furthermore, this thickness-dependent modulation, where the interplay between classical surface energy (which favors a smooth $(111)$ surface) and the quantum size effect (QSE) energy (which favors bi-layer stability), gives rise to rich phenomena in atomic processes of film growth as well as kinetic phenomena \cite{Li2006,Altfeder1997,Ozer2005,Upton2004,Yeh2000,Chan2006,Han2008}.
\nolinebreak
\begin{figure}
\includegraphics[width=3.35 in]{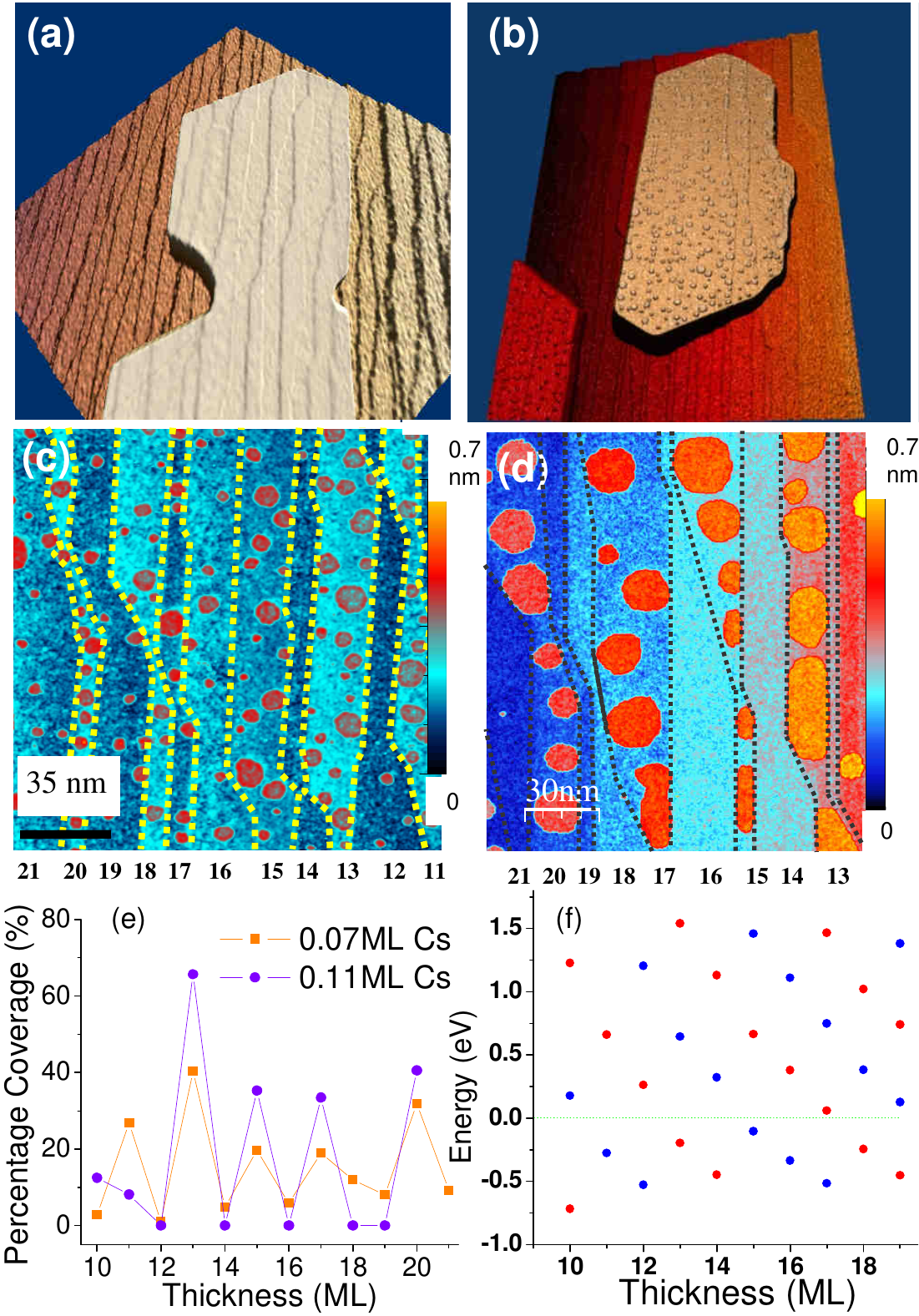}
\caption{\label{fig1}Surface alloy driven nano-island formation: (a) 3D STM example of a flat-top Pb mesa template before Cs deposition ($750 \times  750$  nm$^{2}$); $V_{sample} = +150$ mV, $I = 50$ pA. (b) Similar Pb mesa after $0.07 \pm 0.02$ ML Cs deposition. Nano-islands on the mesa top are the signature of Pb/Cs surface alloy formation. $V_{sample} = -1.5$ V. (c) Bi-layer distribution of nano-islands for $0.07 \pm 0.02$ ML Cs coverage. Dashed lines indicate QWS boundaries resulting from underlying substrate steps. $V_{sample} = -1.5$ V, $I = 30$ pA. (d) Ratio of nano-island coverage to total terrace area for each thickness. $0.07 \pm 0.02$ ML Cs square (orange/bright), $0.11 \pm 0.03$ ML Cs circle (purple/dark). Coverage error $\approx 5 \%$. (e) $0.11 \pm 0.03$ ML Cs; $V_{sample} = -0.5$ V, $I = 65$ pA. (f) Thickness-dependent QWS energy values as determined from STS.  All images taken at $T = 6$ K.}
\end{figure}

In this paper, we report on observations of a set of intricately related atomic processes as we use a combined experimental and theoretical approach to explore the surface alloying and restructuring of quantum metal films.  We use scanning tunneling microscopy and spectroscopy (STM/STS) aided by first-principles simulations within density functional theory (DFT) to demonstrate that the adsorption of a minute amount of Cs on a Pb mesa grown in the quantum regime can induce dramatic morphological changes of the mesa, characterized by the appearance of populous monatomic-layer-high Pb nano-islands on top of the mesa. The edges of the Pb nano-islands are decorated with Cs adatoms, and the nano-islands preferentially nucleate and grow on the quantum mechanically unstable regions of the mesa. Our DFT studies further show that the Pb atoms forming the nano-islands were expelled by the adsorbed Cs atoms via a kinetically accessible place exchange process when the Cs atoms alloyed into the top layer of the Pb mesa.

\section{Experimental and Computational Methods}

All experiments were carried out using a home-built low temperature 4 K STM at a base pressure $< 1 \times 10^{-10}$ Torr. PtIr tips and single crystal Ir tips were used for all experiments. STM measurements were performed between $T = 5.2-6.5$ K where indicated. Vicinal Si$(111)$ samples ($1.1^{\circ}$ miscut toward $[\overline{1} \overline{1} 2]$) utilized were \textit{n}-doped (Sb) with a resistivity of $0.022-0.06$  $\Omega$-cm \cite{Lin1998}.  Pb was deposited on Si$(111)-7 \times 7$ from a thermal evaporator with typical growth rates of 0.5 ML per minute at room temperature, resulting in flat-top mesas \cite{Okamoto2002,Jiang2004,Li2006}.  Cs was deposited from a resistively heated SAES getter source and calibrated relative to the $1 \times 1$ unit cell of Pb, where 1 ML Cs is equivalent to 1 Cs atom sitting at each Pb surface lattice site. Pressure during Cs deposition was kept below $ 5 \times 10^{-10}$ Torr. For Pb flat-top mesas, Cs was deposited at a sample temperature between 100 and 120 K and further annealed to 165 K - 230 K.\\
\nolinebreak
\indent First-principles calculations are based on density functional theory as implemented in the VASP code \cite{Kresse1996}. We used ultrasoft-pseudopotentials \cite{Kresse1994,Vanderbilt1990} with plane-wave basis sets and the generalized gradient approximation (PW91-GGA) \cite{Perdew1992}. For Pb and Cs, the outmost \emph{s} and \emph{p} electrons are treated as valence electrons. STM images were simulated using the Tersoff-Hamman approach \cite{Tersoff1983,Tersoff1985}.

\section{Experimental Results}

Shown in Fig. \ref{fig1}(a) is a Pb mesa template with a flat-top geometry grown on a Si$(111)$ substrate following a room temperature growth procedure \cite{Okamoto2002,Jiang2004,Li2006}. Due to the underlying Si surface steps, such a flat top mesa contains regions with different local film thicknesses, including both quantum mechanically stable and unstable regions. Upon sub-monolayer Cs adsorption, the flat-top Pb mesa undergoes a profound surface re-structuring; this is exemplified by the formation of monolayer height nano-islands such as those shown in Fig. \ref{fig1}(b) after $0.07 \pm 0.02$ ML Cs deposition. Interestingly, the distribution of such nano-islands shows strong oscillations as a function of film thickness.
\nolinebreak
\begin{figure*}
\includegraphics[width=5.5 in]{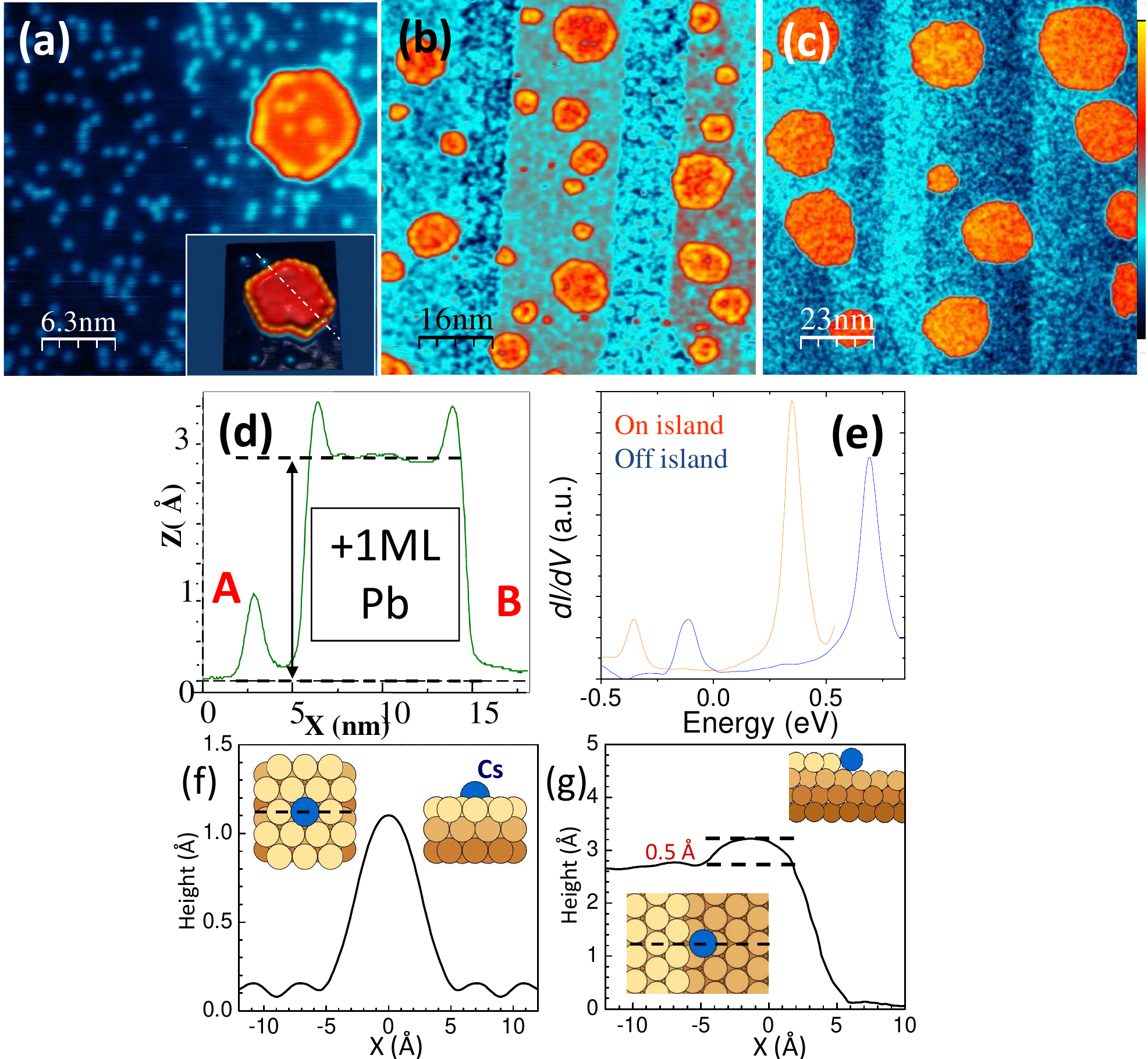}
\caption{\label{fig2}Coverage-dependent evolution of nano-islands and step decorations: (a) 0.024 ML Cs; $V_{sample} = +1$ V; (inset) 3D rendered STM image of a nano-island at $0.01 \pm 0.005$ ML Cs coverage; $V_{sample} = +1$ V. (b) $0.07 \pm 0.02$ ML Cs; $V_{sample} = -1.5$ V, (c) $0.11 \pm 0.03$ ML Cs; $V_{sample} = -500$ mV. (d) Cross-section profile of the nano-island in inset (dashed line). (e) STS taken on/off the nano-island indicated in the inset. The QWS on the nano-island (orange) top correlate with a more quantum favorable thickness compared to the adjacent terrace (blue). STS was taken using lock-in amplification with $V_{mod} = 5$ mV, $f_{mod} = 1.7$ kHz.  Simulated STM images of (f) a Cs atom bound to a surface substitutional site and (g) a Cs atom bound to the lower step edge of a Pb film. Both images were simulated at $V_{sample} = +1$ V. Insets indicate top and side views of the atomic structures.}
\end{figure*}

Fig. \ref{fig1}(c)-(d) show STM characterization of a spatial distribution of nano-islands for $0.07 \pm 0.02$ and $0.11 \pm 0.03$ ML Cs coverage respectively. The imposed dashed lines indicate QWS boundaries resulting from underlying substrate steps. As shown in Fig. \ref{fig1}(d), the nano-island density versus thickness exhibits strong oscillations with a bi-layer periodicity. This oscillatory behavior is observed for both coverages, with maxima appearing on odd layers between 11 and 18 ML where the highest occupied QWS is closer to $E_{F}$.  This thickness dependence of nano-island density correlates very well with the location of the QWS on a Pb mesa prior to Cs-deposition (Fig. \ref{fig1}(f)): that is to say, higher island densities occur at thicknesses where the QWS is closer to $E_{F}$ (commonly referred to as ``quantum unstable'' thicknesses) \cite{Wei2002,Eom2006,Jia2006}. This suggests that nano-island formation is strongly influenced by the underlying electronic structure of the film. As shown in Fig. \ref{fig1}(d), increasing the coverage of Cs (0.11 ML) increases the overall size of these nano-islands. Typical nano-island diameters are on the order of $10-80$ nm, with the average diameter being heavily dependent on the surface alloy density (Cs dosing). Furthermore, as the typical diameter of an individual nano-island becomes larger than the underlying terrace width, a typical nano-island elongates parallel to the substrate step becoming ellipsoidal in shape rather than crossing the neighboring boundary (substrate step). This indicates that it is energetically favorable for a nano-island to remain on quantum unstable layers.
\nolinebreak
\begin{figure}
\includegraphics[width=3.35 in]{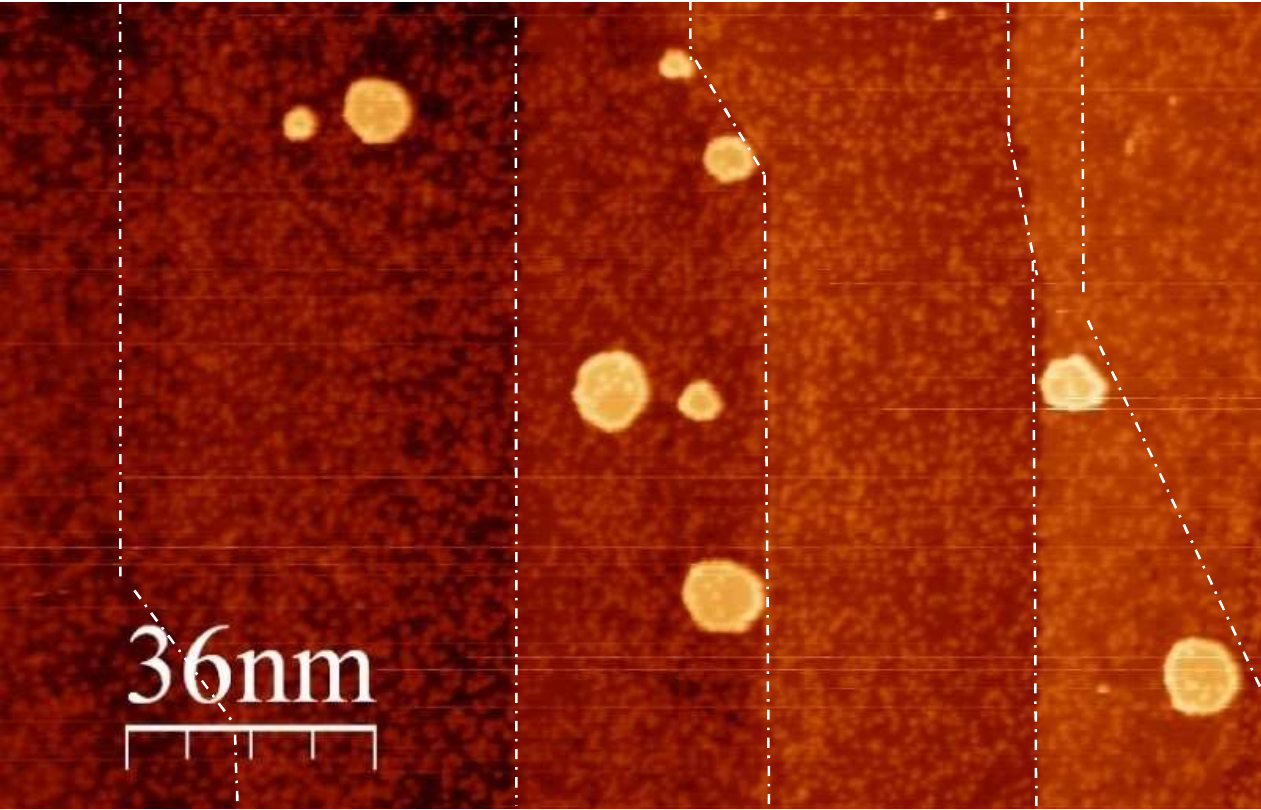}
\caption{\label{fig3}Mass conservation between Pb/Cs atoms: Mesa surface after 0.024 ML Cs deposition;  $V_{sample} = -1.7$ V, $T = 6$ K. White dashed lines were added to delineate absolute thickness changes. The number of atoms which makes up the displayed Pb nano-islands is $\approx 8000$. Nano-islands were approximated as hexagonal. The number of Cs atoms, calculated from the density of Cs per unit area, is $\approx 6000$.}
\end{figure}

The evolution of nano-island shape as a function of Cs coverage (from 0.01 to 0.11 ML) is further illustrated in Fig. \ref{fig2}. At a very low coverage of $0.01 \pm 0.005$ ML (inset of Fig. \ref{fig2}(a)), one can see that circular nano-islands are decorated with single Cs atoms along the nano-island edge. As shown by the line profile in Fig. \ref{fig2}(d), the nano-island height (before incorporating large amounts of Cs) is exactly one monolayer of Pb, which confirms that the nano-islands are Pb in composition. Moreover, STS measurements indicate that Cs-decorated nano-islands have a LDOS identical to +1 ML of Pb further demonstrating that these nano-islands are Pb in composition (Fig. \ref{fig2}(e)).  The apparent topographic height across the isolated Cs atom near the nano-island is 0.9 \AA. This value is too small for an isolated Cs atom lying on top of the surface. Indeed, simulated STM images indicate that a Cs atom on the surface would have a topographic height of more than 3.0 \AA\ while a substitutional Cs atom will result in an apparent topographic height of about 1.0 \AA\ (Fig. \ref{fig2}(f)). We thus conclude that isolated Cs atoms are incorporated into substitutional sites. On the other hand, the apparent topographic height of the Cs atoms at the nano-island edge is about 0.6 \AA\ above the nano-island surface. This value suggests that Cs atoms bind to the lower (ascending) step edge of Pb nano-islands. Simulated STM images from first-principles calculations with a positive sample bias of 1.0 eV confirm that Cs atoms preferentially bind to the lower step edge and have a height profile of $\approx 0.5$ \AA\ relative to the nano-island terrace, which is in good agreement with experimental observations (Fig. \ref{fig2}(g)).\\
\nolinebreak
\indent We noted that the formation of Pb nano-islands is intricately linked to this edge-decoration of Cs impurities. The Cs ring decoration is clearly observed for low coverages (0.01 and 0.024 ML) and remains observable for 0.07 ML coverage, except that the nano-island center now incorporates more Cs. As more Cs atoms are incorporated, the apparent topographic height of the nano-island center is raised, resulting in the disappearance of the contrast of the edge ring.  The direct correlation between island formation and preferential Cs step binding indicates that Cs impurities must play a key role in this morphological transformation.\\
\nolinebreak
\indent The Pb adatoms responsible for nano-island formation are generated by the place-exchange process between Cs and Pb atoms which produces the surface alloy \cite{Nielsen1993,Bardi1994,Besenbacher1998}. It is unlikely that any mesa surface has a large number of vacancy sites before Cs exposure required for the observed Cs intermixing. Numerical analysis reveals that the number of Pb atoms which makes up the nano-islands is roughly equal to the number of surrounding Cs atoms embedded in the mesa surface. This strongly suggests that the Pb atoms responsible for nano-island nucleation come from the Pb adatoms which are popped out of the first layer of the mesa during the place exchange process. No nano-islands were observed without a background distribution of Cs substitutional atoms on the mesa surface.

This source of Pb adatoms is unlike the case where Pb adatoms diffuse up the sidewall of the mesa forming edge-islands as demonstrated in ref \onlinecite{Okamoto2002,Jiang2004,Li2006}.  If edge-island diffusion is responsible for island nucleation, then there should be a strong correlation between the island coverage and edge-island nucleation. However, STM images show nearly uniform island distributions along the direction parallel to the underlying substrate step with negligible edge-island nucleation regardless of island density or Cs exposure.
\nolinebreak
\begin{figure}
\includegraphics[width=3.35 in]{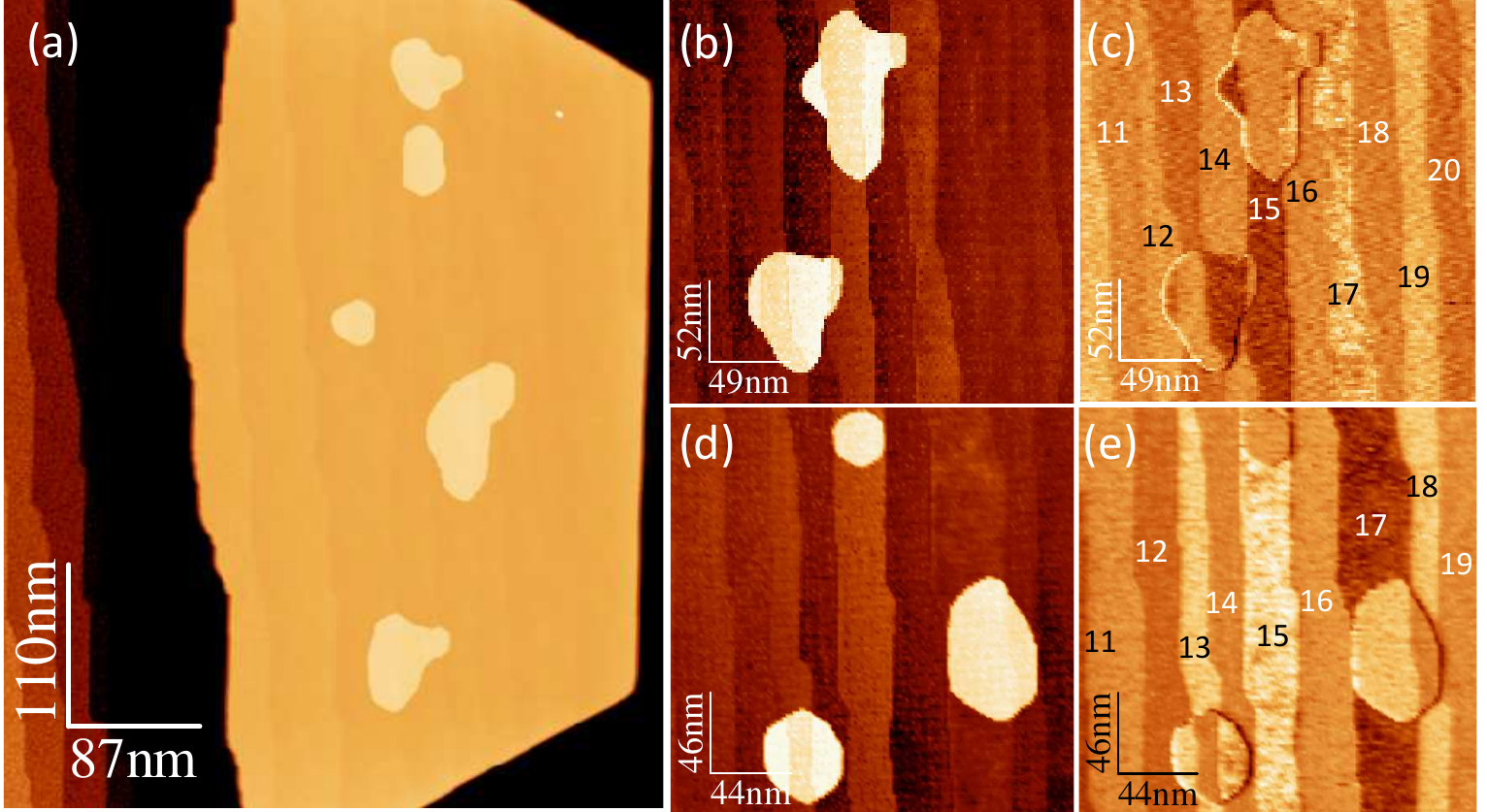}
\caption{\label{fig4}(a) 0.1ML of Pb deposited on an impurity-free Pb mesa and annealed to 160 K $V_{sample} = +1$ V.  Pb islands form on top of a mesa after Pb deposition, (b) Topography; (c) $dI/dV$, $V_{sample} = +30$ mV (d) Topograhy; (e) $dI/dV$, $V_{sample} = -30$ mV.  Numbers indicated the absolute thickness of the mesa.}
\end{figure}

It is important to characterize the role of both kinetics and thermodynamics in the formation of the observed Cs-decorated Pb nano-islands and the specific role Cs atoms play in modifying this energetic balance. To elucidate the importance of Cs in stabilizing Pb nano-islands preferentially on quantum unstable layers, 0.1 ML of Pb was deposited on an impurity-free Pb mesa and was annealed to 160 K identical to the procedure utilized for Cs deposition.  Fig \ref{fig4} illustrates the resultant mesa after additional Pb deposition. Monolayer-high Pb islands can be seen distributed across the mesa surface. Differential conductance maps ($dI/dV$), illustrated in Fig. \ref{fig4}(c),(e), demonstrate that these Pb islands remain +1 ML in height, but can easily span 2 to 3 different layers which include both quantum stable and unstable layers. This is in stark contrast to the Cs-decorated nano-islands which show strong preference to quantum unstable layers and show a circular geometry.  This contrasting behavior suggests that the Cs step decoration plays an important role in the formation of Pb nano-islands which preferably reside on the quantum unstable layers.\\
\nolinebreak
\indent We next discuss the role of kinetic processes such as thickness-dependent surface diffusion, island coarsening, or mesa edge barrier fluctuations on the observed bi- layer island density oscillation \cite{Ma2006,Chan2006}.  Surface diffusion of Pb adatoms is relatively slower on quantum stable regions \cite{Chan2006}, therefore a higher island density is expected to be observed on quantum stable regions \cite{Ma2006}.  Furthermore, bi-layer oscillations in Pb surface diffusion are small in energy suggesting that the observed nano-island distribution oscillations should be heavily temperature dependent. However, the observed thickness-dependent nano-island distributions are very robust as this phenomenon is insensitive to annealing parameters and rather sensitive to the Cs coverage. Embedded Cs atoms, which generate Pb adatoms, are uniformly distributed across the mesa surface independent of thickness, indicating that Pb adatoms responsible for island nucleation non-preferentially originate from both quantum stable and unstable regions.  Nevertheless, observed Cs-decorated Pb nano-islands are distributed with a strong thickness-dependence. At the utilized preparation temperatures, where kinetically-driven place exchange is active, both Pb and Cs adatoms are highly mobile allowing Pb adatoms to nucleate and Cs adatoms to step decorate these nucleated nano-islands. No nano-islands were observed without accompanying Cs step decorations. Moreover, all Cs-decorated nano-islands are nearly circular in shape indicating that they are at a local equilibrium.  Thus, while kinetics definitely play a role in island nucleation and shape relaxation, the resulting bi-layer oscillation of nano-island distribution is a quasi-equilibrium phenomenon at the preparation temperature \cite{note10}.
\nolinebreak
\begin{figure}
\includegraphics[width=3.35 in]{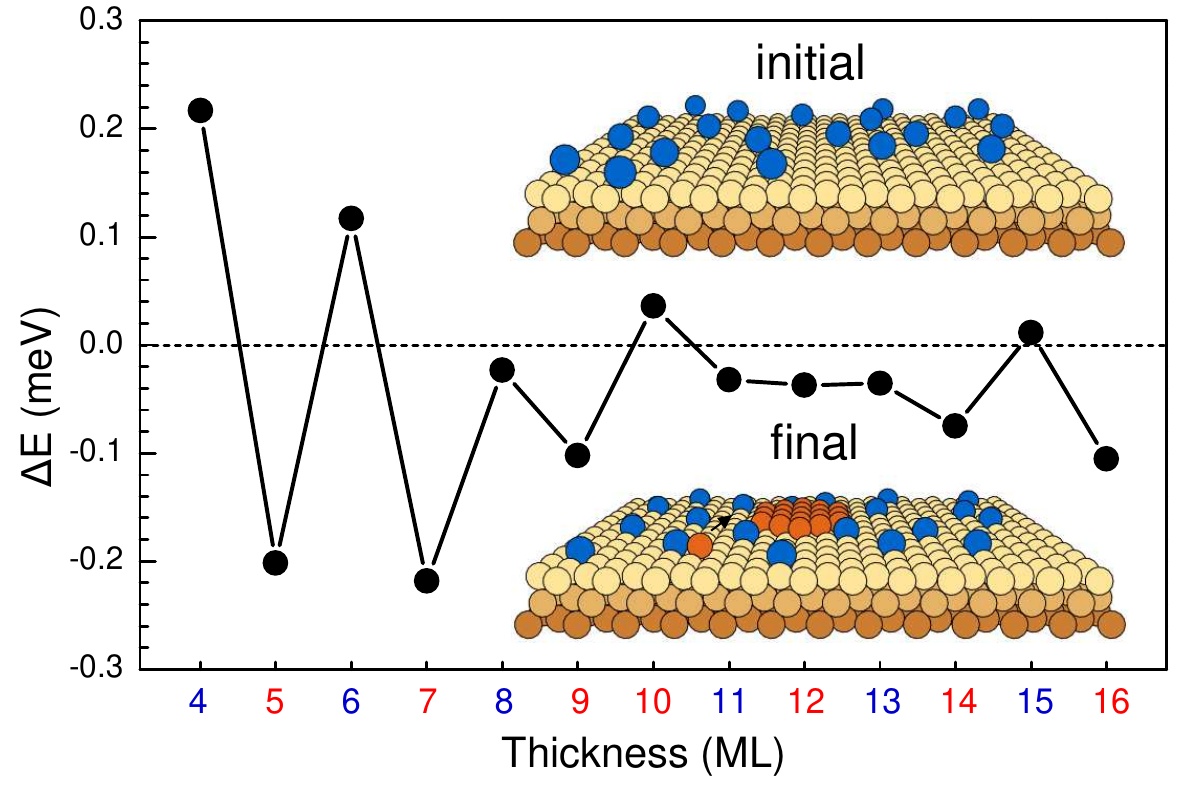}
\caption{\label{fig5}Energetic balance ($\Delta E$, as defined in the text) between the formation of a Cs/Pb surface alloy and a Pb nano-island. Schematic illustrations of the initial and final structures are shown in the inset. Blue and orange balls represent Cs and Pb atoms, respectively.}
\end{figure}

\section{Theoretical Modeling}

To understand why nano-islands form preferentially on the quantum mechanically unstable layers, we consider two configurations, as shown in Fig. \ref{fig5}: (i) the initial state, where isolated Cs atoms are on top of a flat \emph{N}-layer Pb film, and (ii) the final state, where the Cs atoms replace the first-layer Pb atoms and the popped out Pb atoms nucleate to form a nano-island. By neglecting the step formation energy of the nano-island, the total energy difference $\Delta E$ per Pb atom in the two configurations can be estimated as $[E_{Cs-embedded}(N)+E_{Pb-film}(N+1)]-[E_{Pb-film}(N)+E_{Cs-top}(N)]$, where $E_{Pb-film}(N)$ represents the total energy of an $N$-layer Pb film and $E_{Cs-embedded}$, $E_{Cs-top}$ represent the total energy of an $N$-layer Pb film with a Cs atom embedded in the surface and on the top of the surface, respectively. The calculated values of $\Delta E$, from first principles, are plotted in Fig. \ref{fig5} for film thicknesses ranging from 4 to 16 ML.  This calculation, which does not take into account the underlying interface, can produce differences in absolute thickness between experiment and calculation \cite{Jia2006}. Positive (negative) values of $\Delta E$ correspond to energy cost (gain) for the nano-island formation. The results demonstrate that on quantum mechanically unstable layers the configuration with embedded Cs atoms accompanied by nano-island formation is energetically more favorable than a flat-top geometry with a distribution of Cs adatoms.

We also find that Cs island step decorations further promote island stability by lowering the overall cost of the step formation energy. First-principles calculations indicate that Cs adatoms repel each other on the Pb(111) surface. Therefore, the step energy lowered by the Cs step decoration can be estimated by comparing the total energy of a stepped Pb film with a Cs atom binding at the lower step edge with a Cs atom on a flat terrace far away from the step. The calculated energy difference for $\{111\}$ and $\{100\}$ faceted steps is around 200 meV \emph{per Cs atom}. For a clean Pb(111) surface, the step formation energy was calculated to be around 88 meV \emph{per step unit} \cite{Feibelman2000}. In our system, the average inter-atomic spacing per Cs step atom is $ \approx  8.5$ \AA\, which is slightly less than 3 times the step unit length of Pb$(111)$ (3.5 \AA). Therefore, step formation energy cost is greatly lowered by the Cs step decorations, which enhances island stability and plays a significant role during the initial island formation. However, the step energy is proportional to island radius $r$, while the surface energy is proportional to $r^2$. For sufficiently large islands as observed in our experiments, the surface energy term is dominant and therefore responsible for the observed bi-layer island distribution.

\section{Summary}

 The combined theoretical and experimental investigations reveal an intricate interplay between surface alloy formation and quantum size effects, which results in the restructuring of Pb flat-top mesa surfaces in the quantum growth regime. More specifically, by introducing a minute concentration of alkali adsorbates on the surface of a Pb mesa, the energetic balance between the classical and QSE energetic terms can be controllably altered, leading to the formation of monolayer-high Cs step-decorated Pb nano-islands predominately on quantum unstable regions. Such systems, where electron confinement can be controlled by growth conditions, offer interesting platforms for elucidating the underlying physical origins of the enhanced chemical reactivity of bimetallic alloys formed at the 2D surface, potentially offering a new route towards quantum design of catalysts.

\section{Acknowledgments}
We thank Prof. Z. Q. Qiu at UC Berkeley and Prof. Y. Wu at Fudan University for fruitful discussions.  This work was funded by NSF-FRG grant number DMR-0606485, NSF-IGERT grant number DGE-0549417, Alexander von Humboldt-Foundation, and the DMSE program and grant number DE-FG02-05ER46209 of USDOE, the Welch Foundation, and the Texas Advanced Research Program.  The calculations were performed at NERSC of DOE.

\bibliography{nano-islands}

\begin{thebibliography}{10}%
\makeatletter
\providecommand \@ifxundefined [1]{%
 \ifx #1\undefined \expandafter \@firstoftwo
 \else \expandafter \@secondoftwo
\fi
}%
\providecommand \@ifnum [1]{%
 \ifnum #1\expandafter \@firstoftwo
 \else \expandafter \@secondoftwo
\fi
}%
\providecommand \enquote [1]{``#1''}%
\providecommand \bibnamefont  [1]{#1}%
\providecommand \bibfnamefont [1]{#1}%
\providecommand \citenamefont [1]{#1}%
\providecommand\href[0]{\@sanitize\@href}%
\providecommand\@href[1]{\endgroup\@@startlink{#1}\endgroup\@@href}%
\providecommand\@@href[1]{#1\@@endlink}%
\providecommand \@sanitize [0]{\begingroup\catcode`\&12\catcode`\#12\relax}%
\@ifxundefined \pdfoutput {\@firstoftwo}{%
 \@ifnum{\z@=\pdfoutput}{\@firstoftwo}{\@secondoftwo}%
}{%
 \providecommand\@@startlink[1]{\leavevmode\special{html:<a href="#1">}}%
 \providecommand\@@endlink[0]{\special{html:</a>}}%
}{%
 \providecommand\@@startlink[1]{%
  \leavevmode
  \pdfstartlink
   attr{/Border[0 0 1 ]/H/I/C[0 1 1]}%
   user{/Subtype/Link/A<</Type/Action/S/URI/URI(#1)>>}%
  \relax
 }%
 \providecommand\@@endlink[0]{\pdfendlink}%
}%
\providecommand \url  [0]{\begingroup\@sanitize \@url }%
\providecommand \@url [1]{\endgroup\@href {#1}{\urlprefix}}%
\providecommand \urlprefix [0]{URL }%
\providecommand \Eprint[0]{\href }%
\@ifxundefined \urlstyle {%
  \providecommand \doi [1]{doi:\discretionary{}{}{}#1}%
}{%
  \providecommand \doi [0]{doi:\discretionary{}{}{}\begingroup
  \urlstyle{rm}\Url }%
}%
\providecommand \doibase [0]{http://dx.doi.org/}%
\providecommand \Doi[1]{\href{\doibase#1}}%
\providecommand \bibAnnote [3]{%
  \BibitemShut{#1}%
  \begin{quotation}\noindent
    \textsc{Key:}\ #2\\\textsc{Annotation:}\ #3%
  \end{quotation}%
}%
\providecommand \bibAnnoteFile [2]{%
  \IfFileExists{#2}{\bibAnnote {#1} {#2} {\input{#2}}}{}%
}%
\providecommand \typeout [0]{\immediate \write \m@ne }%
\providecommand \selectlanguage [0]{\@gobble}%
\providecommand \bibinfo [0]{\@secondoftwo}%
\providecommand \bibfield [0]{\@secondoftwo}%
\providecommand \translation [1]{[#1]}%
\providecommand \BibitemOpen[0]{}%
\providecommand \bibitemStop [0]{}%
\providecommand \bibitemNoStop [0]{.\EOS\space}%
\providecommand \EOS [0]{\spacefactor3000\relax}%
\providecommand \BibitemShut [1]{\csname bibitem#1\endcsname}%
\bibitem{Brune1998}%
  \BibitemOpen
  \bibfield{author}{%
  \bibinfo {author} {\bibfnamefont{H.}~\bibnamefont{Brune}}, \bibinfo {author}
  {\bibfnamefont{M.}~\bibnamefont{Giovannini}}, \bibinfo {author}
  {\bibfnamefont{K.}~\bibnamefont{Bromann}},\ and\ \bibinfo {author}
  {\bibfnamefont{K.}~\bibnamefont{Kern}},\ }%
  \bibfield{journal}{%
  \bibinfo {journal} {Nature}\ }%
  \textbf{\bibinfo {volume} {394}},\ \bibinfo {pages} {451} (\bibinfo {year}
  {1998})%
  \bibAnnoteFile{NoStop}{Brune1998}%
\bibitem{Ernst1998}%
  \BibitemOpen
  \bibfield{author}{%
  \bibinfo {author} {\bibfnamefont{H.~J.}\ \bibnamefont{Ernst}}, \bibinfo
  {author} {\bibfnamefont{F.}~\bibnamefont{Charra}},\ and\ \bibinfo {author}
  {\bibfnamefont{L.}~\bibnamefont{Douillard}},\ }%
  \bibfield{journal}{%
  \bibinfo {journal} {Science}\ }%
  \textbf{\bibinfo {volume} {279}},\ \bibinfo {pages} {679} (\bibinfo {year}
  {1998})%
  \bibAnnoteFile{NoStop}{Ernst1998}%
\bibitem{Peale1992}%
  \BibitemOpen
  \bibfield{author}{%
  \bibinfo {author} {\bibfnamefont{D.~R.}\ \bibnamefont{Peale}}\ and\ \bibinfo
  {author} {\bibfnamefont{B.~H.}\ \bibnamefont{Cooper}},\ }%
  \bibfield{journal}{%
  \bibinfo {journal} {J. Vac. Sci. Technol. A}\ }%
  \textbf{\bibinfo {volume} {10}},\ \bibinfo {pages} {2210} (\bibinfo {year}
  {1992})%
  \bibAnnoteFile{NoStop}{Peale1992}%
\bibitem{Ross1999}%
  \BibitemOpen
  \bibfield{author}{%
  \bibinfo {author} {\bibfnamefont{F.~M.}\ \bibnamefont{Ross}}, \bibinfo
  {author} {\bibfnamefont{R.~M.}\ \bibnamefont{Tromp}},\ and\ \bibinfo {author}
  {\bibfnamefont{M.~C.}\ \bibnamefont{Reuter}},\ }%
  \bibfield{journal}{%
  \bibinfo {journal} {Science}\ }%
  \textbf{\bibinfo {volume} {286}},\ \bibinfo {pages} {1931} (\bibinfo {year}
  {1999})%
  \bibAnnoteFile{NoStop}{Ross1999}%
\bibitem{Stangl2004}%
  \BibitemOpen
  \bibfield{author}{%
  \bibinfo {author} {\bibfnamefont{J.}~\bibnamefont{Stangl}}, \bibinfo {author}
  {\bibfnamefont{V.}~\bibnamefont{Holý}},\ and\ \bibinfo {author}
  {\bibfnamefont{G.}~\bibnamefont{Bauer}},\ }%
  \bibfield{journal}{%
  \bibinfo {journal} {Rev. Mod. Phys.}\ }%
  \textbf{\bibinfo {volume} {76}},\ \bibinfo {pages} {725} (\bibinfo {year}
  {2004})%
  \bibAnnoteFile{NoStop}{Stangl2004}%
\bibitem{Aballe2004}%
  \BibitemOpen
  \bibfield{author}{%
  \bibinfo {author} {\bibfnamefont{L.}~\bibnamefont{Aballe}}, \bibinfo {author}
  {\bibfnamefont{A.}~\bibnamefont{Barinov}}, \bibinfo {author}
  {\bibfnamefont{A.}~\bibnamefont{Locatelli}}, \bibinfo {author}
  {\bibfnamefont{S.}~\bibnamefont{Heun}},\ and\ \bibinfo {author}
  {\bibfnamefont{M.}~\bibnamefont{Kiskinova}},\ }%
  \bibfield{journal}{%
  \bibinfo {journal} {Phys.Rev. Lett.}\ }%
  \textbf{\bibinfo {volume} {93}},\ \bibinfo {pages} {196103} (\bibinfo {year}
  {2004})%
  \bibAnnoteFile{NoStop}{Aballe2004}%
\bibitem{Eom2006}%
  \BibitemOpen
  \bibfield{author}{%
  \bibinfo {author} {\bibfnamefont{D.}~\bibnamefont{Eom}}, \bibinfo {author}
  {\bibfnamefont{S.}~\bibnamefont{Qin}}, \bibinfo {author}
  {\bibfnamefont{M.~Y.}\ \bibnamefont{Chou}},\ and\ \bibinfo {author}
  {\bibfnamefont{C.~K.}\ \bibnamefont{Shih}},\ }%
  \bibfield{journal}{%
  \bibinfo {journal} {Phys. Rev. Lett.}\ }%
  \textbf{\bibinfo {volume} {96}},\ \bibinfo {pages} {027005} (\bibinfo {year}
  {2006})%
  \bibAnnoteFile{NoStop}{Eom2006}%
\bibitem{Evans1993}%
  \BibitemOpen
  \bibfield{author}{%
  \bibinfo {author} {\bibfnamefont{D.~A.}\ \bibnamefont{Evans}}, \bibinfo
  {author} {\bibfnamefont{M.}~\bibnamefont{Alonso}}, \bibinfo {author}
  {\bibfnamefont{R.}~\bibnamefont{Cimino}},\ and\ \bibinfo {author}
  {\bibfnamefont{K.}~\bibnamefont{Horn}},\ }%
  \bibfield{journal}{%
  \bibinfo {journal} {Phys. Rev. Lett.}\ }%
  \textbf{\bibinfo {volume} {70}},\ \bibinfo {pages} {3483} (\bibinfo {year}
  {1993})%
  \bibAnnoteFile{NoStop}{Evans1993}%
\bibitem{Guo2004}%
  \BibitemOpen
  \bibfield{author}{%
  \bibinfo {author} {\bibfnamefont{Y.}~\bibnamefont{Guo}} \emph{et~al.},\ }%
  \bibfield{journal}{%
  \bibinfo {journal} {Science}\ }%
  \textbf{\bibinfo {volume} {306}},\ \bibinfo {pages} {1915} (\bibinfo {year}
  {2004})%
  \bibAnnoteFile{NoStop}{Guo2004}%
\bibitem{Hinch1989}%
  \BibitemOpen
  \bibfield{author}{%
  \bibinfo {author} {\bibfnamefont{B.~J.}\ \bibnamefont{Hinch}}, \bibinfo
  {author} {\bibfnamefont{C.}~\bibnamefont{Koziol}}, \bibinfo {author}
  {\bibfnamefont{J.~P.}\ \bibnamefont{Toennies}},\ and\ \bibinfo {author}
  {\bibfnamefont{G.}~\bibnamefont{Zhang}},\ }%
  \bibfield{journal}{%
  \bibinfo {journal} {Europhys. Lett.}\ }%
  \textbf{\bibinfo {volume} {10}},\ \bibinfo {pages} {341} (\bibinfo {year}
  {1989})%
  \bibAnnoteFile{NoStop}{Hinch1989}%
\bibitem{Miller1988}%
  \BibitemOpen
  \bibfield{author}{%
  \bibinfo {author} {\bibfnamefont{T.}~\bibnamefont{Miller}}, \bibinfo {author}
  {\bibfnamefont{A.}~\bibnamefont{Samsavar}}, \bibinfo {author}
  {\bibfnamefont{G.~E.}\ \bibnamefont{Franklin}},\ and\ \bibinfo {author}
  {\bibfnamefont{T.~C.}\ \bibnamefont{Chiang}},\ }%
  \bibfield{journal}{%
  \bibinfo {journal} {Phys. Rev. Lett.}\ }%
  \textbf{\bibinfo {volume} {61}},\ \bibinfo {pages} {1404} (\bibinfo {year}
  {1988})%
  \bibAnnoteFile{NoStop}{Miller1988}%
\bibitem{Ozer2007}%
  \BibitemOpen
  \bibfield{author}{%
  \bibinfo {author} {\bibfnamefont{M.~M.}\ \bibnamefont{\"{O}zer}}, \bibinfo
  {author} {\bibfnamefont{Y.}~\bibnamefont{Jia}}, \bibinfo {author}
  {\bibfnamefont{Z.~Y.}\ \bibnamefont{Zhang}}, \bibinfo {author}
  {\bibfnamefont{J.~R.}\ \bibnamefont{Thompson}},\ and\ \bibinfo {author}
  {\bibfnamefont{H.~H.}\ \bibnamefont{Weitering}},\ }%
  \bibfield{journal}{%
  \bibinfo {journal} {Science}\ }%
  \textbf{\bibinfo {volume} {316}},\ \bibinfo {pages} {1594} (\bibinfo {year}
  {2007})%
  \bibAnnoteFile{NoStop}{Ozer2007}%
\bibitem{Smith1996}%
  \BibitemOpen
  \bibfield{author}{%
  \bibinfo {author} {\bibfnamefont{A.~R.}\ \bibnamefont{Smith}}, \bibinfo
  {author} {\bibfnamefont{K.~J.}\ \bibnamefont{Chao}}, \bibinfo {author}
  {\bibfnamefont{Q.}~\bibnamefont{Niu}},\ and\ \bibinfo {author}
  {\bibfnamefont{C.~K.}\ \bibnamefont{Shih}},\ }%
  \bibfield{journal}{%
  \bibinfo {journal} {Science}\ }%
  \textbf{\bibinfo {volume} {273}},\ \bibinfo {pages} {226} (\bibinfo {year}
  {1996})%
  \bibAnnoteFile{NoStop}{Smith1996}%
\bibitem{Zhang1998}%
  \BibitemOpen
  \bibfield{author}{%
  \bibinfo {author} {\bibfnamefont{Z.~Y.}\ \bibnamefont{Zhang}}, \bibinfo
  {author} {\bibfnamefont{Q.}~\bibnamefont{Niu}},\ and\ \bibinfo {author}
  {\bibfnamefont{C.~K.}\ \bibnamefont{Shih}},\ }%
  \bibfield{journal}{%
  \bibinfo {journal} {Phys. Rev. Lett.}\ }%
  \textbf{\bibinfo {volume} {80}},\ \bibinfo {pages} {5381} (\bibinfo {year}
  {1998})%
  \bibAnnoteFile{NoStop}{Zhang1998}%
\bibitem{Wei2002}%
  \BibitemOpen
  \bibfield{author}{%
  \bibinfo {author} {\bibfnamefont{C.~M.}\ \bibnamefont{Wei}}\ and\ \bibinfo
  {author} {\bibfnamefont{M.~Y.}\ \bibnamefont{Chou}},\ }%
  \bibfield{journal}{%
  \bibinfo {journal} {Phys. Rev. B}\ }%
  \textbf{\bibinfo {volume} {66}},\ \bibinfo {pages} {233408} (\bibinfo {year}
  {2002})%
  \bibAnnoteFile{NoStop}{Wei2002}%
\bibitem{Jia2006}%
  \BibitemOpen
  \bibfield{author}{%
  \bibinfo {author} {\bibfnamefont{Y.}~\bibnamefont{Jia}}, \bibinfo {author}
  {\bibfnamefont{B.}~\bibnamefont{Wu}}, \bibinfo {author}
  {\bibfnamefont{H.~H.}\ \bibnamefont{Weitering}},\ and\ \bibinfo {author}
  {\bibfnamefont{Z.~Y.}\ \bibnamefont{Zhang}},\ }%
  \bibfield{journal}{%
  \bibinfo {journal} {Phys. Rev. B}\ }%
  \textbf{\bibinfo {volume} {74}},\ \bibinfo {pages} {035433} (\bibinfo {year}
  {2006})%
  \bibAnnoteFile{NoStop}{Jia2006}%
\bibitem{Ma2006}%
  \BibitemOpen
  \bibfield{author}{%
  \bibinfo {author} {\bibfnamefont{L.-Y.}\ \bibnamefont{Ma}} \emph{et~al.},\ }%
  \bibfield{journal}{%
  \bibinfo {journal} {Phys. Rev. Lett.}\ }%
  \textbf{\bibinfo {volume} {97}},\ \bibinfo {pages} {266102} (\bibinfo {year}
  {2006})%
  \bibAnnoteFile{NoStop}{Ma2006}%
\bibitem{Ma2007}%
  \BibitemOpen
  \bibfield{author}{%
  \bibinfo {author} {\bibfnamefont{X.~C.}\ \bibnamefont{Ma}}, \bibinfo {author}
  {\bibfnamefont{P.}~\bibnamefont{Jiang}}, \bibinfo {author}
  {\bibfnamefont{Y.}~\bibnamefont{Qi}}, \bibinfo {author}
  {\bibfnamefont{J.~F.}\ \bibnamefont{Jia}}, \bibinfo {author}
  {\bibfnamefont{Y.}~\bibnamefont{Yang}}, \bibinfo {author}
  {\bibfnamefont{W.~H.}\ \bibnamefont{Duan}}, \bibinfo {author}
  {\bibfnamefont{W.~X.}\ \bibnamefont{Li}}, \bibinfo {author}
  {\bibfnamefont{X.}~\bibnamefont{Bao}}, \bibinfo {author}
  {\bibfnamefont{S.~B.}\ \bibnamefont{Zhang}},\ and\ \bibinfo {author}
  {\bibfnamefont{Q.~K.}\ \bibnamefont{Xue}},\ }%
  \bibfield{journal}{%
  \bibinfo {journal} {Proc. Natl. Acad. Sci. USA}\ }%
  \textbf{\bibinfo {volume} {104}},\ \bibinfo {pages} {9204} (\bibinfo {year}
  {2007})%
  \bibAnnoteFile{NoStop}{Ma2007}%
\bibitem{Li2006}%
  \BibitemOpen
  \bibfield{author}{%
  \bibinfo {author} {\bibfnamefont{S.~C.}\ \bibnamefont{Li}}, \bibinfo {author}
  {\bibfnamefont{X.~C.}\ \bibnamefont{Ma}}, \bibinfo {author}
  {\bibfnamefont{J.~F.}\ \bibnamefont{Jia}}, \bibinfo {author}
  {\bibfnamefont{Y.~F.}\ \bibnamefont{Zhang}}, \bibinfo {author}
  {\bibfnamefont{D.~M.}\ \bibnamefont{Chen}}, \bibinfo {author}
  {\bibfnamefont{Q.}~\bibnamefont{Niu}}, \bibinfo {author}
  {\bibfnamefont{F.}~\bibnamefont{Liu}}, \bibinfo {author}
  {\bibfnamefont{P.~S.}\ \bibnamefont{Weiss}},\ and\ \bibinfo {author}
  {\bibfnamefont{Q.~K.}\ \bibnamefont{Xue}},\ }%
  \bibfield{journal}{%
  \bibinfo {journal} {Phys. Rev. B}\ }%
  \textbf{\bibinfo {volume} {74}},\ \bibinfo {pages} {075410} (\bibinfo {year}
  {2006})%
  \bibAnnoteFile{NoStop}{Li2006}%
\bibitem{Altfeder1997}%
  \BibitemOpen
  \bibfield{author}{%
  \bibinfo {author} {\bibfnamefont{I.~B.}\ \bibnamefont{Altfeder}}, \bibinfo
  {author} {\bibfnamefont{K.~A.}\ \bibnamefont{Matveev}},\ and\ \bibinfo
  {author} {\bibfnamefont{D.~M.}\ \bibnamefont{Chen}},\ }%
  \bibfield{journal}{%
  \bibinfo {journal} {Phys. Rev. Lett.}\ }%
  \textbf{\bibinfo {volume} {78}},\ \bibinfo {pages} {2815} (\bibinfo {year}
  {1997})%
  \bibAnnoteFile{NoStop}{Altfeder1997}%
\bibitem{Ozer2005}%
  \BibitemOpen
  \bibfield{author}{%
  \bibinfo {author} {\bibfnamefont{M.~M.}\ \bibnamefont{\"{O}zer}}, \bibinfo
  {author} {\bibfnamefont{Y.}~\bibnamefont{Jia}}, \bibinfo {author}
  {\bibfnamefont{B.}~\bibnamefont{Wu}}, \bibinfo {author}
  {\bibfnamefont{Z.~Y.}\ \bibnamefont{Zhang}},\ and\ \bibinfo {author}
  {\bibfnamefont{H.~H.}\ \bibnamefont{Weitering}},\ }%
  \bibfield{journal}{%
  \bibinfo {journal} {Phys. Rev. B}\ }%
  \textbf{\bibinfo {volume} {72}},\ \bibinfo {pages} {235427} (\bibinfo {year}
  {2005})%
  \bibAnnoteFile{NoStop}{Ozer2005}%
\bibitem{Upton2004}%
  \BibitemOpen
  \bibfield{author}{%
  \bibinfo {author} {\bibfnamefont{M.~H.}\ \bibnamefont{Upton}}, \bibinfo
  {author} {\bibfnamefont{C.~M.}\ \bibnamefont{Wei}}, \bibinfo {author}
  {\bibfnamefont{M.~Y.}\ \bibnamefont{Chou}}, \bibinfo {author}
  {\bibfnamefont{T.}~\bibnamefont{Miller}},\ and\ \bibinfo {author}
  {\bibfnamefont{T.~C.}\ \bibnamefont{Chiang}},\ }%
  \bibfield{journal}{%
  \bibinfo {journal} {Phys. Rev. Lett.}\ }%
  \textbf{\bibinfo {volume} {93}},\ \bibinfo {pages} {026802} (\bibinfo {year}
  {2004})%
  \bibAnnoteFile{NoStop}{Upton2004}%
\bibitem{Yeh2000}%
  \BibitemOpen
  \bibfield{author}{%
  \bibinfo {author} {\bibfnamefont{V.}~\bibnamefont{Yeh}}, \bibinfo {author}
  {\bibfnamefont{L.}~\bibnamefont{Berbil-Bautista}}, \bibinfo {author}
  {\bibfnamefont{C.~Z.}\ \bibnamefont{Wang}}, \bibinfo {author}
  {\bibfnamefont{K.~M.}\ \bibnamefont{Ho}},\ and\ \bibinfo {author}
  {\bibfnamefont{M.~C.}\ \bibnamefont{Tringides}},\ }%
  \bibfield{journal}{%
  \bibinfo {journal} {Phys. Rev. Lett.}\ }%
  \textbf{\bibinfo {volume} {85}},\ \bibinfo {pages} {5158} (\bibinfo {year}
  {2000})%
  \bibAnnoteFile{NoStop}{Yeh2000}%
\bibitem{Chan2006}%
  \BibitemOpen
  \bibfield{author}{%
  \bibinfo {author} {\bibfnamefont{T.~L.}\ \bibnamefont{Chan}}, \bibinfo
  {author} {\bibfnamefont{C.}~\bibnamefont{Wang}}, \bibinfo {author}
  {\bibfnamefont{M.}~\bibnamefont{Hupalo}}, \bibinfo {author}
  {\bibfnamefont{M.}~\bibnamefont{Tringides}},\ and\ \bibinfo {author}
  {\bibfnamefont{K.}~\bibnamefont{Ho}},\ }%
  \bibfield{journal}{%
  \bibinfo {journal} {Phys. Rev. Lett.}\ }%
  \textbf{\bibinfo {volume} {96}},\ \bibinfo {pages} {226102} (\bibinfo {year}
  {2006})%
  \bibAnnoteFile{NoStop}{Chan2006}%
\bibitem{Han2008}%
  \BibitemOpen
  \bibfield{author}{%
  \bibinfo {author} {\bibfnamefont{Y.}~\bibnamefont{Han}}, \bibinfo {author}
  {\bibfnamefont{M.}~\bibnamefont{Hupalo}}, \bibinfo {author}
  {\bibfnamefont{M.~C.}\ \bibnamefont{Tringides}},\ and\ \bibinfo {author}
  {\bibfnamefont{F.}~\bibnamefont{Liu}},\ }%
  \bibfield{journal}{%
  \bibinfo {journal} {Surf. Sci.}\ }%
  \textbf{\bibinfo {volume} {602}},\ \bibinfo {pages} {62} (\bibinfo {year}
  {2008})%
  \bibAnnoteFile{NoStop}{Han2008}%
\bibitem{Lin1998}%
  \BibitemOpen
  \bibfield{author}{%
  \bibinfo {author} {\bibfnamefont{J.-L.}\ \bibnamefont{Lin}}, \bibinfo
  {author} {\bibfnamefont{D.~Y.}\ \bibnamefont{Petrovykh}}, \bibinfo {author}
  {\bibfnamefont{J.}~\bibnamefont{Viernow}}, \bibinfo {author}
  {\bibfnamefont{F.~K.}\ \bibnamefont{Men}}, \bibinfo {author}
  {\bibfnamefont{D.~J.}\ \bibnamefont{Seo}},\ and\ \bibinfo {author}
  {\bibfnamefont{F.~J.}\ \bibnamefont{Himpsel}},\ }%
  \bibfield{journal}{%
  \bibinfo {journal} {J. Appl. Phys.}\ }%
  \textbf{\bibinfo {volume} {84}},\ \bibinfo {pages} {255} (\bibinfo {year}
  {1998})%
  \bibAnnoteFile{NoStop}{Lin1998}%
\bibitem{Okamoto2002}%
  \BibitemOpen
  \bibfield{author}{%
  \bibinfo {author} {\bibfnamefont{H.}~\bibnamefont{Okamoto}}, \bibinfo
  {author} {\bibfnamefont{D.~M.}\ \bibnamefont{Chen}},\ and\ \bibinfo {author}
  {\bibfnamefont{T.}~\bibnamefont{Yamada}},\ }%
  \bibfield{journal}{%
  \bibinfo {journal} {Phys. Rev. Lett.}\ }%
  \textbf{\bibinfo {volume} {89}},\ \bibinfo {pages} {256101} (\bibinfo {year}
  {2002})%
  \bibAnnoteFile{NoStop}{Okamoto2002}%
\bibitem{Jiang2004}%
  \BibitemOpen
  \bibfield{author}{%
  \bibinfo {author} {\bibfnamefont{C.~S.}\ \bibnamefont{Jiang}}, \bibinfo
  {author} {\bibfnamefont{S.~C.}\ \bibnamefont{Li}}, \bibinfo {author}
  {\bibfnamefont{H.~B.}\ \bibnamefont{Yu}}, \bibinfo {author}
  {\bibfnamefont{D.}~\bibnamefont{Eom}}, \bibinfo {author}
  {\bibfnamefont{X.~D.}\ \bibnamefont{Wang}}, \bibinfo {author}
  {\bibfnamefont{P.}~\bibnamefont{Ebert}}, \bibinfo {author}
  {\bibfnamefont{J.~F.}\ \bibnamefont{Jia}}, \bibinfo {author}
  {\bibfnamefont{Q.~K.}\ \bibnamefont{Xue}},\ and\ \bibinfo {author}
  {\bibfnamefont{C.~K.}\ \bibnamefont{Shih}},\ }%
  \bibfield{journal}{%
  \bibinfo {journal} {Phys. Rev. Lett.}\ }%
  \textbf{\bibinfo {volume} {92}},\ \bibinfo {pages} {106104} (\bibinfo {year}
  {2004})%
  \bibAnnoteFile{NoStop}{Jiang2004}%
\bibitem{Kresse1996}%
  \BibitemOpen
  \bibfield{author}{%
  \bibinfo {author} {\bibfnamefont{G.}~\bibnamefont{Kresse}}\ and\ \bibinfo
  {author} {\bibfnamefont{J.}~\bibnamefont{Furthmuller}},\ }%
  \bibfield{journal}{%
  \bibinfo {journal} {Phys. Rev. B}\ }%
  \textbf{\bibinfo {volume} {54}},\ \bibinfo {pages} {11169} (\bibinfo {year}
  {1996})%
  \bibAnnoteFile{NoStop}{Kresse1996}%
\bibitem{Kresse1994}%
  \BibitemOpen
  \bibfield{author}{%
  \bibinfo {author} {\bibfnamefont{G.}~\bibnamefont{Kresse}}\ and\ \bibinfo
  {author} {\bibfnamefont{J.}~\bibnamefont{Hafner}},\ }%
  \bibfield{journal}{%
  \bibinfo {journal} {J. Phys.: Condens. Matter}\ }%
  \textbf{\bibinfo {volume} {6}},\ \bibinfo {pages} {8245} (\bibinfo {year}
  {1994})%
  \bibAnnoteFile{NoStop}{Kresse1994}%
\bibitem{Vanderbilt1990}%
  \BibitemOpen
  \bibfield{author}{%
  \bibinfo {author} {\bibfnamefont{D.}~\bibnamefont{Vanderbilt}},\ }%
  \bibfield{journal}{%
  \bibinfo {journal} {Phys. Rev. B}\ }%
  \textbf{\bibinfo {volume} {41}},\ \bibinfo {pages} {7892} (\bibinfo {year}
  {1990})%
  \bibAnnoteFile{NoStop}{Vanderbilt1990}%
\bibitem{Perdew1992}%
  \BibitemOpen
  \bibfield{author}{%
  \bibinfo {author} {\bibfnamefont{J.~P.}\ \bibnamefont{Perdew}}\ and\ \bibinfo
  {author} {\bibfnamefont{Y.}~\bibnamefont{Wang}},\ }%
  \bibfield{journal}{%
  \bibinfo {journal} {Phys. Rev. B}\ }%
  \textbf{\bibinfo {volume} {45}},\ \bibinfo {pages} {13244} (\bibinfo {year}
  {1992})%
  \bibAnnoteFile{NoStop}{Perdew1992}%
\bibitem{Tersoff1983}%
  \BibitemOpen
  \bibfield{author}{%
  \bibinfo {author} {\bibfnamefont{J.}~\bibnamefont{Tersoff}}\ and\ \bibinfo
  {author} {\bibfnamefont{D.~R.}\ \bibnamefont{Hamann}},\ }%
  \bibfield{journal}{%
  \Doi{10.1103/PhysRevLett.50.1998}{\bibinfo {journal} {Phys. Rev. Lett.}}\ }%
  \textbf{\bibinfo {volume} {50}},\ \bibinfo {pages} {1998} (\bibinfo {month}
  {Jun}\ \bibinfo {year} {1983})%
  \bibAnnoteFile{NoStop}{Tersoff1983}%
\bibitem{Tersoff1985}%
  \BibitemOpen
  \bibfield{author}{%
  \bibinfo {author} {\bibfnamefont{J.}~\bibnamefont{Tersoff}}\ and\ \bibinfo
  {author} {\bibfnamefont{D.~R.}\ \bibnamefont{Hamann}},\ }%
  \bibfield{journal}{%
  \Doi{10.1103/PhysRevB.31.805}{\bibinfo {journal} {Phys. Rev. B}}\ }%
  \textbf{\bibinfo {volume} {31}},\ \bibinfo {pages} {805} (\bibinfo {month}
  {Jan}\ \bibinfo {year} {1985})%
  \bibAnnoteFile{NoStop}{Tersoff1985}%
\bibitem{Nielsen1993}%
  \BibitemOpen
  \bibfield{author}{%
  \bibinfo {author} {\bibfnamefont{L.}~\bibnamefont{Pleth~Nielsen}}, \bibinfo
  {author} {\bibfnamefont{F.}~\bibnamefont{Besenbacher}}, \bibinfo {author}
  {\bibfnamefont{I.}~\bibnamefont{Stensgaard}}, \bibinfo {author}
  {\bibfnamefont{E.}~\bibnamefont{Laegsgaard}}, \bibinfo {author}
  {\bibfnamefont{C.}~\bibnamefont{Engdahl}}, \bibinfo {author}
  {\bibfnamefont{P.}~\bibnamefont{Stoltze}}, \bibinfo {author}
  {\bibfnamefont{K.~W.}\ \bibnamefont{Jacobsen}},\ and\ \bibinfo {author}
  {\bibfnamefont{J.~K.}\ \bibnamefont{N\o{}rskov}},\ }%
  \bibfield{journal}{%
  \Doi{10.1103/PhysRevLett.71.754}{\bibinfo {journal} {Phys. Rev. Lett.}}\ }%
  \textbf{\bibinfo {volume} {71}},\ \bibinfo {pages} {754} (\bibinfo {month}
  {Aug}\ \bibinfo {year} {1993})%
  \bibAnnoteFile{NoStop}{Nielsen1993}%
\bibitem{Bardi1994}%
  \BibitemOpen
  \bibfield{author}{%
  \bibinfo {author} {\bibfnamefont{U.}~\bibnamefont{Bardi}},\ }%
  \bibfield{journal}{%
  \bibinfo {journal} {Rep. Prog. Phys.}\ }%
  \textbf{\bibinfo {volume} {57}},\ \bibinfo {pages} {939} (\bibinfo {year}
  {1994})%
  \bibAnnoteFile{NoStop}{Bardi1994}%
\bibitem{Besenbacher1998}%
  \BibitemOpen
  \bibfield{author}{%
  \bibinfo {author} {\bibfnamefont{F.}~\bibnamefont{Besenbacher}}, \bibinfo
  {author} {\bibfnamefont{I.}~\bibnamefont{Chorkendorff}}, \bibinfo {author}
  {\bibfnamefont{B.~S.}\ \bibnamefont{Clausen}}, \bibinfo {author}
  {\bibfnamefont{B.}~\bibnamefont{Hammer}}, \bibinfo {author}
  {\bibfnamefont{A.~M.}\ \bibnamefont{Molenbroek}}, \bibinfo {author}
  {\bibfnamefont{J.~K.}\ \bibnamefont{N\o{}rskov}},\ and\ \bibinfo {author}
  {\bibfnamefont{I.}~\bibnamefont{Stensgaard}},\ }%
  \bibfield{journal}{%
  \bibinfo {journal} {Science}\ }%
  \textbf{\bibinfo {volume} {279}},\ \bibinfo {pages} {1913} (\bibinfo {year}
  {1998})%
  \bibAnnoteFile{NoStop}{Besenbacher1998}%
\bibitem{note10}%
  \BibitemOpen
  \bibinfo {note} {For the case of 0.1 ML Pb deposited on an impurity-free Pb
  mesa, when the annealing temperature is raised to 230 K, the observed Pb
  islands indeed reside primarily on quantum unstable regions implying that
  quasi-equilbrium is reached at a higher temperature than the case for
  Cs-decorated nano-islands.}%
  \bibAnnoteFile{Stop}{note10}%
\bibitem{Feibelman2000}%
  \BibitemOpen
  \bibfield{author}{%
  \bibinfo {author} {\bibfnamefont{P.~J.}\ \bibnamefont{Feibelman}},\ }%
  \bibfield{journal}{%
  \bibinfo {journal} {Phys. Rev. B}\ }%
  \textbf{\bibinfo {volume} {62}},\ \bibinfo {pages} {17020} (\bibinfo {year}
  {2000})%
  \bibAnnoteFile{NoStop}{Feibelman2000}%
\end{thebibliography}%

\end{document}